\documentclass[preprint,amsmath,amssymb,aps,nofootinbib,superscriptaddress]{revtex4-2}

\usepackage{graphicx}
\usepackage{dcolumn}
\usepackage{bm}
\usepackage{xcolor}
\usepackage[normalem]{ulem}
\usepackage{hyperref}

\begin{document}


\title{Covalency, correlations, and inter-layer interactions governing the magnetic and electronic structure of Mn$_3$Si$_2$Te$_6$}

\author{Chiara Bigi}
\affiliation{SUPA, School of Physics and Astronomy, University of St Andrews, St Andrews KY16 9SS, UK}

\author{Lei Qiao}
\affiliation{Physics Department, International Center of Quantum and Molecular Structures, Materials Genome Institute, State Key Laboratory of Advanced Special Steel, Shanghai Key Laboratory of High Temperature Superconductors, Shanghai University, Shanghai 200444, China}
\affiliation{Consiglio Nazionale delle Ricerche (CNR-SPIN), Unità di Ricerca presso Terzi c/o Università “G. D'Annunzio”, 66100 Chieti, Italy}

\author{Chao Liu}
\affiliation{Physics Department, International Center of Quantum and Molecular Structures, Materials Genome Institute, State Key Laboratory of Advanced Special Steel, Shanghai Key Laboratory of High Temperature Superconductors, Shanghai University, Shanghai 200444, China}
\affiliation{Consiglio Nazionale delle Ricerche (CNR-SPIN), Unità di Ricerca presso Terzi c/o Università “G. D'Annunzio”, 66100 Chieti, Italy}

\author{Paolo Barone}
\affiliation{Consiglio Nazionale delle Ricerche CNR-SPIN, Area della Ricerca di Tor Vergata, Via del Fosso del Cavaliere 100, I-00133 Rome, Italy}

\author{Monica  \surname{Ciomaga Hatnean}}
\altaffiliation{Current address: Paul Scherrer Institut, Forschungsstrasse 111, 5232 Villigen PSI, Switzerland}
\affiliation {Department of Physics, University of Warwick, Coventry, CV4 7AL, United Kingdom}

\author{Gesa-R. Siemann}
\affiliation{SUPA, School of Physics and Astronomy, University of St Andrews, St Andrews KY16 9SS, UK}

\author{Barat Achinuq}
\affiliation{Department of Physics, Clarendon Laboratory, University of Oxford, Oxford, OX1~3PU, United Kingdom}

\author{Daniel Alexander Mayoh}
\affiliation {Department of Physics, University of Warwick, Coventry, CV4 7AL, United Kingdom}

\author{Giovanni Vinai}
\affiliation{Istituto Officina dei Materiali (IOM)-CNR, Laboratorio TASC, Area Science Park, S.S.14, Km 163.5, 34149 Trieste, Italy}

\author{Vincent Polewczyk}
\affiliation{Istituto Officina dei Materiali (IOM)-CNR, Laboratorio TASC, Area Science Park, S.S.14, Km 163.5, 34149 Trieste, Italy}

\author{Deepak Dagur} 
\affiliation{Istituto Officina dei Materiali (IOM)-CNR, Laboratorio TASC, Area Science Park, S.S.14, Km 163.5, 34149 Trieste, Italy}

\author{Federico Mazzola} 
\affiliation{Istituto Officina dei Materiali (IOM)-CNR, Laboratorio TASC, Area Science Park, S.S.14, Km 163.5, 34149 Trieste, Italy}

\author{Peter Bencok}
\affiliation{Diamond Light Source, Harwell Science and Innovation Campus, Didcot, OX11~0DE, United Kingdom}

\author{Thorsten Hesjedal}
\affiliation{Department of Physics, Clarendon Laboratory, University of Oxford, Oxford, OX1~3PU, United Kingdom}

\author{Gerrit \surname{van der Laan}}
\affiliation{Diamond Light Source, Harwell Science and Innovation Campus, Didcot, OX11~0DE, United Kingdom}

\author{Wei Ren}
\affiliation{Physics Department, International Center of Quantum and Molecular Structures, Materials Genome Institute, State Key Laboratory of Advanced Special Steel, Shanghai Key Laboratory of High Temperature Superconductors, Shanghai University, Shanghai 200444, China}

\author{Geetha Balakrishnan}
\affiliation {Department of Physics, University of Warwick, Coventry, CV4 7AL, United Kingdom}

\author{Silvia Picozzi}
\email{silvia.picozzi@spin.cnr.it}
\affiliation{Consiglio Nazionale delle Ricerche (CNR-SPIN), Unità di Ricerca presso Terzi c/o Università “G. D'Annunzio”, 66100 Chieti, Italy}

\author{Phil D. C. King}
\email{pdk6@st-andrews.ac.uk}
\affiliation{SUPA, School of Physics and Astronomy, University of St Andrews, St Andrews KY16 9SS, UK}

\date{\today}

\begin{abstract}
Mn$_3$Si$_2$Te$_6$ is a rare example of a layered ferrimagnet. It has recently been shown to host a colossal angular magnetoresistance as the spin orientation is rotated from the in- to out-of-plane direction, proposed to be underpinned by a topological nodal-line degeneracy in its electronic structure. Nonetheless, the origins of its ferrimagnetic structure remain controversial, while its experimental electronic structure, and the role of correlations in shaping this, are little explored to date. Here, we combine x-ray and photoemission-based spectroscopies with first-principles calculations, to probe the elemental-selective electronic structure and magnetic order in Mn$_3$Si$_2$Te$_6$. Through these, we identify a marked Mn-Te hybridisation, which weakens the electronic correlations and enhances the magnetic anisotropy. We demonstrate how this strengthens the magnetic frustration in Mn$_3$Si$_2$Te$_6$, which is key to stabilising its ferrimagnetic order, and find a crucial role of both exchange interactions extending beyond nearest-neighbours and anti-symmetric exchange in dictating its ordering temperature. Together, our results demonstrate a powerful methodology of using experimental electronic structure probes to constrain the parameter space for first-principles calculations of magnetic materials, and through this approach, reveal a pivotal role played by covalency in stabilising the ferrimagnetic order in Mn$_3$Si$_2$Te$_6$.
\end{abstract}

\maketitle
\section{Introduction}
While three-dimensional magnets are common-place, long range order is strictly forbidden to occur in one-dimensional systems \cite{PRL1966Mermin}. Layered magnetic materials present a novel environment in which to study the critical dimensionality between these two extremes, with finite inter-plane coupling expected to have a strong influence on the development of long-range order, magnetic anisotropy, and the role of fluctuations \cite{Nature2018Burch}. In this respect, Mn$_3$Si$_2$Te$_6$ (MST) is a particularly intriguing compound. It forms in the $P\overline{3}1c$ space group (No.~163, Fig.~\ref{fig:fig1}(a,b)) \cite{JMMM1981Rimet,JSSC1986Vincent}, containing two inequivalent Mn sites. The Mn$_{(1)}$ atoms sit at the centre of edge-sharing MnTe$_6$ octahedra. The resulting Mn atoms would form a triangular lattice. However, $1/3$ of the sites are occupied by a Si-Si dimer, leaving the Mn$_{(1)}$ atoms in a honeycomb configuration (Fig.~\ref{fig:fig1}(b)), akin to the Cr sites of the layered van der Waals magnet Cr$_2$Ge$_2$Te$_6$ (CGT)~\cite{PRL2019Zhang,PRB2019Suzuki,PRB2020Watson}. Unlike CGT, however, extra magnetic ions (Mn$_{(2)}$) are situated between the Mn$_{(1)}$ layers, filling the van der Waals gap. These self-intercalated Mn$_{(2)}$ sites form a triangular lattice, with one Mn$_{(2)}$ atop every second Mn$_{(1)}$ site, providing a bridging link between the Mn$_{(1)}$ layers and establishing a more three-dimensional structure. Nevertheless, the samples cleave easily with a standard top-post method, exposing a flat and uniform surface.

The Mn ions have been reported to develop a long-range magnetic order below $\sim75$~K \cite{PRB2018Liu}, with the Mn moments aligned ferromagnetically within each Mn layer, but with antiferromagnetic coupling between neighbouring layers (Fig.~\ref{fig:fig1}(a))~\cite{PRB2017May}. This leads to an overall ferrimagnetic structure~\cite{PRB2018Liu,JAP2021Olmos}. This speaks to a critical role of the bridging Mn$_{(2)}$ sites, leading to markedly different interlayer interaction as compared to ferromagnetic CGT~\cite{PRB2017May,JMMM2020May}. Indeed, the exchange interactions for the first three nearest-neighbours, $J_1$, $J_2$ and $J_3$ (Fig.~\ref{fig:fig1}(b)) are known to be antiferromagnetic \cite{PRB2017May}. This creates competing interactions which, combined with the geometrical arrangement of the Mn-atoms in the MST lattice (side-view of Fig.~\ref{fig:fig1}(b)), has been proposed to result in a high degree of magnetic frustration in this system. 

It remains poorly understood, however, how the overall ferrimagnetic ground state of MST is stabilised. Moreover, while the Mn ions would nominally be expected in a ${2+}$ valence state with a quenched orbital moment, this has recently been questioned, as a huge magnetic anisotropy (up to $13$~T) has been reported~\cite{PRB2021Ni,Nature2021Seo,PRB2022Sala}. The intra- and inter-layer magnetic couplings in MST thus require further exploration, while the underlying electronic structure from which the magnetic order emerges, and the degree to which ligand hybridisation {\it vs.} electronic correlation effects dominate the magnetic coupling, remain almost completely unexplored to date. 

To address this, here we present a combined theoretical and experimental study into the electronic structure and magnetic interactions of MST. Our results point to a significant Mn-Te orbital hybridisation. We show how this promotes additional exchange coupling terms, inclusion of which is essential to obtain an accurate description of the magnetic anisotropies and ordering temperature in MST. Together, our results thus provide key insights on the interplay and influence of dimensionality, anisotropy, covalency, and correlations on the magnetic fluctuations of quasi-layered magnetic materials.

\section{Methods}

\subsection{DFT} 
We performed density functional theory (DFT) simulations, using the Vienna ab initio Simulation Package (VASP) \cite{PRB1996Kresse,CMS1996Kresse}. The generalized gradient approximation (GGA) based on the Perdew-Burke-Ernzerhof (PBE) functional \cite{PRL1996Perdew} was employed to treat the exchange-correlation interaction. We considered the localized 3$d$ electron correlation of Mn atoms by using the GGA+$U$ method \cite{PRB2006Wang,PRB1995Liechtenstein}, with an effective Hubbard $U$ parameter chosen to be $1$~eV, and other effective $U$ choices of $0$, $2$, and $3$~eV also tested and compared. The projector-augmented-wave (PAW) potentials \cite{PRB1994Bloch,PRB1999Kresse} were used to describe the electron-ion interaction. The energy cutoff was selected to be $550$~eV for the plane-wave basis set. The Brillouin zones were sampled using a 6 × 6 × 4 k-grid mesh in the $\Gamma$-centered scheme. The forces convergence threshold on each atom was chosen as $0.005$~eV/\AA~ and the self-consistent calculations were stopped when the energy difference was smaller than $10^{-7}$~eV per atom. We optimized ionic positions with ground state ferrimagnetic magnetic configuration, and experimental lattice constants were used and fixed in all calculations.

\subsection{Monte Carlo Calculations}
A standard Metropolis algorithm has been used for Monte Carlo (MC) simulations, with 10$^5$ MC steps for equilibration and 5$\times 10^5$ MC steps for averaging. Starting from the crystallographic unit cell, comprising four Mn$_{(1)}$ and two Mn$_{(2)}$ sites, we performed calculations on a 16$\times$16$\times$8 supercell with $N_s=12288$ spins. The transition temperature can be estimated from the peak appearing in the temperature evolution of the specific heat per spin, evaluated as
\begin{eqnarray}\label{cv}
C_v &=&\frac{k_\mathrm{B}\beta^2}{N_s}\left[\langle E^2\rangle - \langle E\rangle^2 \right]
\end{eqnarray}
where $E$ is the energy calculated using model Eq. (\ref{eq:model}), $k_B$ is the Boltzmann constant and $\beta=1/k_\mathrm{B}T$, while $\langle ...\rangle$ indicates statistical averages. We also define the ferromagnetic $\bm F$ and ferrimagnetic $\bm f$ order parameters as:
\begin{eqnarray}
{\bm F} &=& \frac{1}{6}\left(\sum_{i=1,4} \bm S_i + \sum_{i=1,2} \bm s_i\right)\nonumber\\
{\bm f} &=& \frac{1}{6}\left(\sum_{i=1,4} \bm S_i - \sum_{i=1,2} \bm s_i\right)
\end{eqnarray}
where $\bm S_i$ and $\bm s_i$ label here the spins on Mn$_{(1)}$ and Mn$_{(2)}$ sites, respectively, within the unit cell, with associated (generalized) susceptibilities for the magnetic order parameters $OP=F,f$ calculated as:
\begin{eqnarray}
\chi^{OP} &=&\beta N_s \left[\langle OP^2\rangle - \langle OP\rangle^2  \right]
\end{eqnarray}
Results of Monte Carlo calculations are summarized in Fig. \ref{fig:figMC}. We used four sets of parameters in our simulations: two sets correspond to the full model parametrization summarized in Tab. \ref{tab_magnetic_iso} and Tab. \ref{tab_magnetic_aniso}, while two sets consist in simplified model with isotropic $J_1^{\mathrm{iso}}, (J_2^{\mathrm{iso}}+J_5^{\mathrm{iso}}),J_3^{\mathrm{iso}}$ exchange interactions and single-ion anisotropy terms. In all cases, we find a transition to an ordered ferrimagnetic phase characterized by ferromagnetic layers of Mn$_{(1)}$ and Mn$_{(2)}$ spins antiferromagnetically aligned and lying in the basal plane. The critical temperature is inversely proportional to the value of U, getting lower as the Hubbard parameter is increased, as expected. On the other hand, the full model parametrization entails a stronger magnetic frustration arising from longer-range antiferromagnetic interactions within the honeycomb layers, that reduces the transition temperature by about 20\% with respect to the simplified model. The range of the estimated critical temperatures settles between 89 and 118 K for $U$ between 2 and 1, respectively.

\subsection{Sample growth and characterisation} Mn$_3$Si$_2$Te$_6$ single-crystals were grown by the chemical vapour transport method using Iodine as a transport agent, as described in Ref.~\cite{JSSC1986Vincent}. X-ray Diffraction (XRD) reported in Supplementary Fig.~S1 show only $(00l)$ Bragg peaks of the expected MST structure, indicating that the facets of the obtained crystals (see inset in Fig.~S1) are parallel to the $ab$-plane. The cell parameters extracted from fitting the XRD data are $a = b = 7.07700(6)$~\AA and $c = 14.25081(2)$~\AA. A Quantum Design Magnetic Property Measurement System was used to measure the magnetization between 1.8 and 300 K in applied fields up to 5 T. A Quantum Design Physical Property Measurement System was used to measure electrical resistivity between 20 and 300 K.

\subsection{X-ray and UV spectroscopies} 
For our spectroscopic measurements, samples were cleaved in ultra-high-vacuum (UHV) using a top-post method. X-ray absorption spectroscopy (XAS) and x-ray magnetic circular dichroism (XMCD) measurements were performed using the electromagnet end station of the I10 beamline at Diamond Light Source, UK. The pressure was better than $10^{-9}$~mbar, and measurements were performed across the Mn $L_{2,3}$ absorption edge in total electron yield (TEY) detection, thus probing $\sim\!6$\,nm depth from the sample surface. Spectra were measured at $10$~K with left- (CL) and right-circularly (CR) polarized x-rays in both normal- and grazing-incidence (i.e., at an angle of $20^\circ$ from the $ab$-plane) geometries (see Supplementary Fig.~S2). An applied field of $1.4$~T collinear with the beam axis was applied to magnetize the sample. The main error source on the XAS/XMCD quantitative data analysis arises from the step-edge background subtraction. To estimate this, each spectrum was analysed using different step-edge background choices (mainly varying the energy range of interest as this was found to affect the background shape the most) and we extracted the standard error deviation from the set of obtained values. The $m_s$ sum rules have been corrected for Mn $jj$-coupling \cite{PRB2005Edmonds} and the temperature dependent XMCD asymmetry has been calculated from the dichroic signal at the $L_3$ absorption peak as
\[A=\frac{I_{\mathrm{CR},L_3}-I_{\mathrm{CL},L_3}}{I_{\mathrm{CR},L_3}+I_{\mathrm{CL},L_3}}.\] The error on the asymmetry has been estimated from the XMCD residual value obtained for several measurements performed well above the transition temperature.

Resonant photoemission spectroscopy (resPES) measurements were performed at the APE-HE beamline (Elettra synchrotron, Italy) at a base pressure lower than $10^{-10}$~mbar, with the sample temperature kept at $\sim107$~K. XAS across the Mn $L_{2,3}$ edge was performed in linear horizontal polarization and in TEY detection to determine the relevant energies for resPES. Measurements were acquired by a Scienta Omicron R3000 hemispherical electron energy analyser. The binding and photon energies were calibrated with the Fermi edge of a gold reference sample, assuming a work function of $4.4$~eV. For better comparison with the experiment, our P-DOS DFT calculations were convolved with a Voigt function to emulate lifetime broadening of $0.01$~eV and an experimental resolution $0.3$~eV.

The ARPES data were acquired at the APE-LE beamline (Elettra synchrotron) with a Scienta DA30 hemispherical electron energy and momentum analyzer, with the samples held at a temperature of $77$~K. The base pressure in the chamber was better than $10^{-10}$~mbar. Energy and angular resolution were better than $40$~meV and $20^\circ$, respectively. 

\section{Results and Discussion}
\subsection{Magnetic order}
\begin{figure*}
\includegraphics[width=0.8\textwidth]{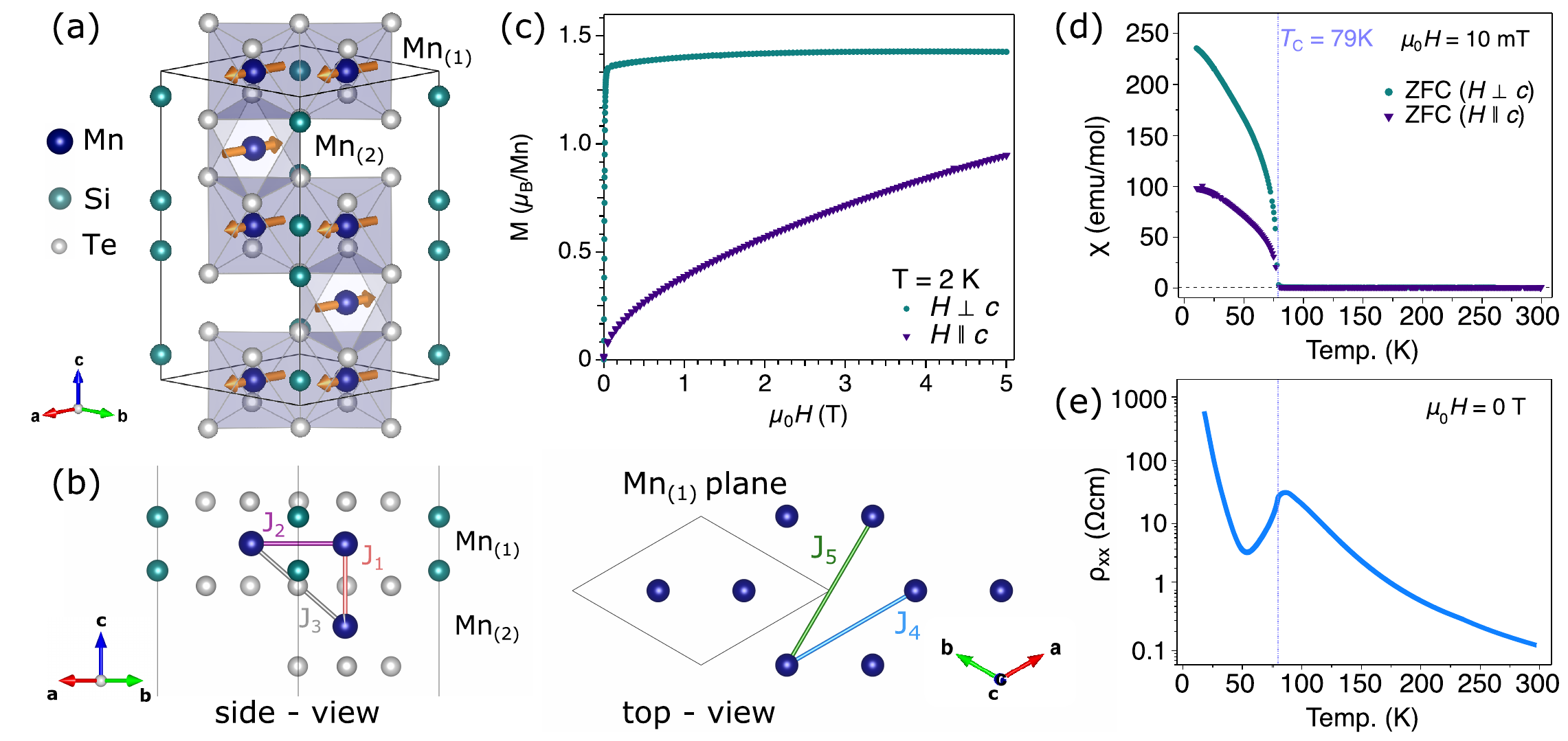}
\caption{\label{fig:fig1} Ferrimagnetism in Mn$_{3}$Si$_{2}$Te$_{6}$. (a) Crystal structure of MST, indicating the ferrimagnetic alignments of Mn moments within the (001) easy $ab$-plane. (b) Side- and Top-views showing the atomic arrangements together with the exchange interactions $J_i$ between the Mn atoms considered in this work. (c) $M$ {\it vs.} $H$ magnetization curves measured at $T=2$~K for $H$~$\parallel$~$c$ and $H$~$\perp$~$c$ geometries. (d) Temperature-dependent dc magnetic susceptibility ($\chi$) as measured in zero-field cooled (ZFC) warming mode in a small field of 10~mT applied along the crystal directions parallel and perpendicular to the $c$-axis. (e) Planar resistivity ($\rho_{xx}$) measured in zero field.}
\end{figure*}

Consistent with prior studies \cite{PRB2017May,PRB2018Liu,APL2020Martinez}, our single-crystal samples (see Supplementary Fig.~S1) exhibit finite net magnetization onsetting below a transition temperature ($T_\mathrm{C}$) of $\sim79$~K (Fig.~\ref{fig:fig1}(d)). $M$ {\it vs.} $H$ magnetization curves performed at $T=2$~K (Fig.~\ref{fig:fig1}(c)) indicate a large magnetic anisotropy with a magnetic hard axis along the $c$ axis: the magnetization fully saturates to $1.42\pm0.02$~$\mu_\mathrm{B}$/Mn in the $ab$-plane ($H\perp{}c$) for less than $2$~T magnetic field, while $5$~T is insufficient to saturate the moment in the $H~\parallel~c$ configuration. Consistent with this large anisotropy, Fig.~\ref{fig:fig1}(d) shows that the magnetic molar susceptibility is more than a factor of two larger for the magnetic field aligned perpendicular {\it vs.} parallel to the $c$-axis. 

\begin{figure}
\includegraphics[width=.8\columnwidth]{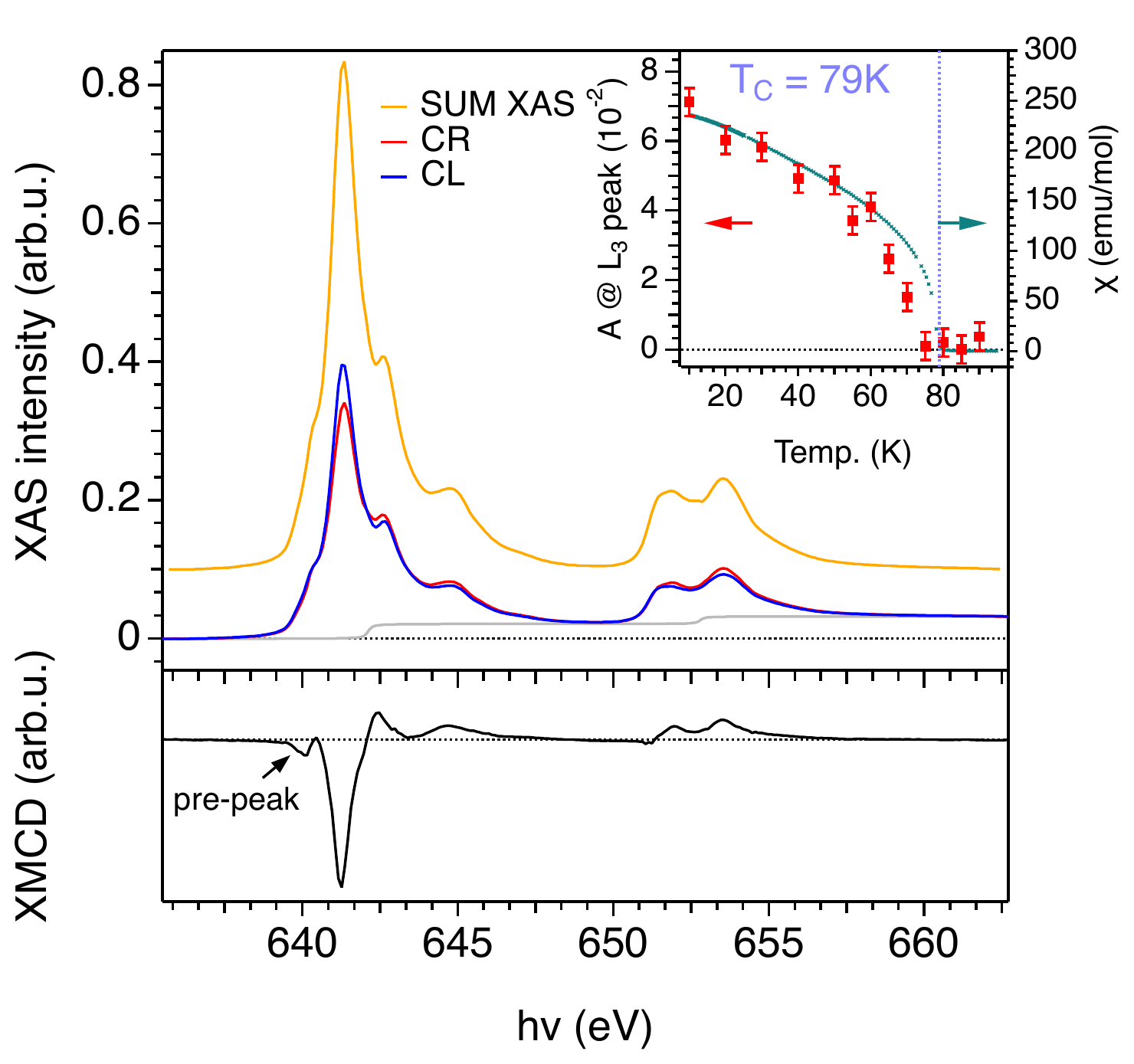}
\caption{\label{fig:fig2} Spectroscopic investigation of the magnetic order in grazing incidence geometry. \emph{Top graph:} Mn $L_{2,3}$ XAS measured at 10~K with an applied field of $1.4$~T for the two opposite x-ray helicities: circular right (CR, red) and circular left (CL, blue). The yellow curve shows the Mn $L_{2,3}$ XAS summed over the CR and CL spectra after the step-edge background (grey) subtraction. \emph{Bottom graph:} x-ray magnetic circular dichroism (XMCD) obtained from the difference between the two absorption spectra. The black arrow denotes the pre-peak feature, which is a fingerprint for Mn hybridisation with the ligand. \emph{Inset:} Temperature-dependent XMCD asymmetry measured at remanence (i.e., in zero applied field as a function of increasing temperature) after first field-cooling the sample. The magnetic susceptibility curve obtained from bulk measurements (green curve) are also shown. The dashed-dotted vertical line marks the magnetic ordering temperature.}
\end{figure}

The simplest picture in which to describe this magnetic order is to consider fully localised Mn spins, with Mn in a 2+ oxidation state, and hence nominal $3d^5$ configuration~\cite{MRSAdv2019Martinez,APL2020Martinez}. Meanwhile, Te$^{2-}$ would be expected to contribute a filled $p^6$ valence band, leading to semiconducting transport (Fig.~\ref{fig:fig1}(e), Refs.~\cite{PRB2017May,PRB2018Liu,PRB2021Ni,PRB2022Wang}). Nonetheless, a pronounced kink in the measured resistivity at the magnetic ordering temperature points to a non-negligible coupling between the magnetism and charge carriers in the system, beyond the simple picture outlined above. To investigate this further, we measured x-ray spectroscopy at the Mn $L_{2,3}$ edge, as shown in Fig.\ref{fig:fig2}. The x-ray absorption measurements (yellow curve in Fig.~\ref{fig:fig2}) shows a spectrum broadly reminiscent of those observed in other nominally Mn$^{2+}$ compounds \cite{APL2000Ohldag,APL2004Edmonds}. The branching ratio $L_3$/($L_3$+$L_2$) exceeds the expected $2/3$ statistical value (i.e., $0.768\pm0.004$), consistent with a nominal Mn high-spin state and with the observed semiconducting transport behavior \cite{PRB1988Thole,PRB1997Durr}. We stress, however, that a picture of fully localised Mn moments is an oversimplification. Hints of this are already detectable in its XAS features, where the characteristic multiplet structure is smeared out and less pronounced than in typical ionic Mn$^{2+}$ compounds \cite{PRB2022Fujii}. This provides a first spectroscopic indication that in fact there is a non-negligible hybridisation of the Mn with the chalcogen ions, a point we return to below. 

Our x-ray absorption data show a clear circular dichroism at low temperature, confirming that the magnetism of MST primarily proceeds via ordering of the Mn spins. The x-ray magnetic circular dichroism (XMCD) signal obtained at $10$~K in a grazing incidence geometry (with a $1.4$~T field applied at $20^\circ$ to the $ab$-plane) is approximately twice as large as for a normal incidence geometry where the field is applied along the $c$-axis, in good agreement with the magnetic susceptibility measured by SQUID (Fig.~\ref{fig:fig1}.(d)). Temperature-dependent XMCD spectra measured in remanence, after the sample was field-cooled to $10$~K in $1.4$~T planar field (inset of Fig.~\ref{fig:fig2}), yield an XMCD asymmetry which closely follows the bulk demagnetization (Fig.~\ref{fig:fig1}(d), reproduced in green in the inset of Fig.~\ref{fig:fig2}). In good agreement with the bulk $T_\mathrm{C}$, the XMCD asymmetry vanishes at $T\approx75$~K becoming comparable to our measurement error (see Fig. S2 in Supplementary Material for the full temperature-dependent XMCD spectra).

Having established that our spectroscopic measurements probe bulk-like magnetic properties in MST, we turn to the insights on the magnetic and electronic structure which they advance. The main spectrum arises from the multiplet structure of the Mn $d^5$ states that are split by the crystal-field interaction. In addition, we observe a distinct pre-peak feature of $(-,+)$ shape, marked by a black arrow in the bottom graph of Fig.\ref{fig:fig2}. Such a pre-edge feature in other Mn-based compounds has been ascribed to transitions from the Mn $2p$ core level to unoccupied $p$-$d$ hybridized valence states~\cite{vanderLaan2010}, and thus indicates hybridization with the Te 5$p$ valence states here. Indeed, the shape of our measured XMCD is similar to that found previously for Ga$_{1-x}$Mn$_x$As, where the ground state of Mn is found as a hybridised state with 16\% $d^4$, 58\% $d^5$, and 26\% $d^6$ character, yielding a $d$-count of $\sim\!5.1$ electrons/Mn atom.\cite{APL2004Edmonds}. To consolidate these findings, we applied quantitative sum-rule analysis to our XMCD data-set, to extract the atom-specific orbital and spin magnetic moments ($m_l$ and $m_s$, respectively) \cite{PRL1992Thole,PRL1993Carra,PRB2005Edmonds}. Our results  indicate a smaller than expected ordered spin moment of $m_s=0.546\pm0.004~\mu_\mathrm{B}/\mathrm{Mn}$ - likely due to a finite out of plane component of the applied field for our experimental geometry, thus not reaching saturation for our available fields. Nonetheless, constitent with prior calculations~\cite{PRB2017May}, we find a small but positive orbital moment of $m_l=0.022\pm0.004~\mu_\mathrm{B}/\mathrm{Mn}$ (in agreememt with the DFT value of 0.027 $\mu_\mathrm{B}$ per Mn), with $m_l/m_s=0.040\pm0.005$, further confirming the deviation from a purely high-spin localised picture of the Mn atoms in this system. The positive sign of this ratio points directly to a $d$ shell which is more than half filled, entirely consistent with our above qualitative assessment of the spectral lineshapes observed. Indeed, an orbital to spin ratio of 0.04 tallies with a $d$-count of $\sim$5.1 electrons/Mn, with the small excess of electrons compared to Mn$^{2+}$ implying the presence of holes in the Te 5$p$ band.

\subsection{Electronic structure}
\begin{figure}
\includegraphics[width=0.7\columnwidth]{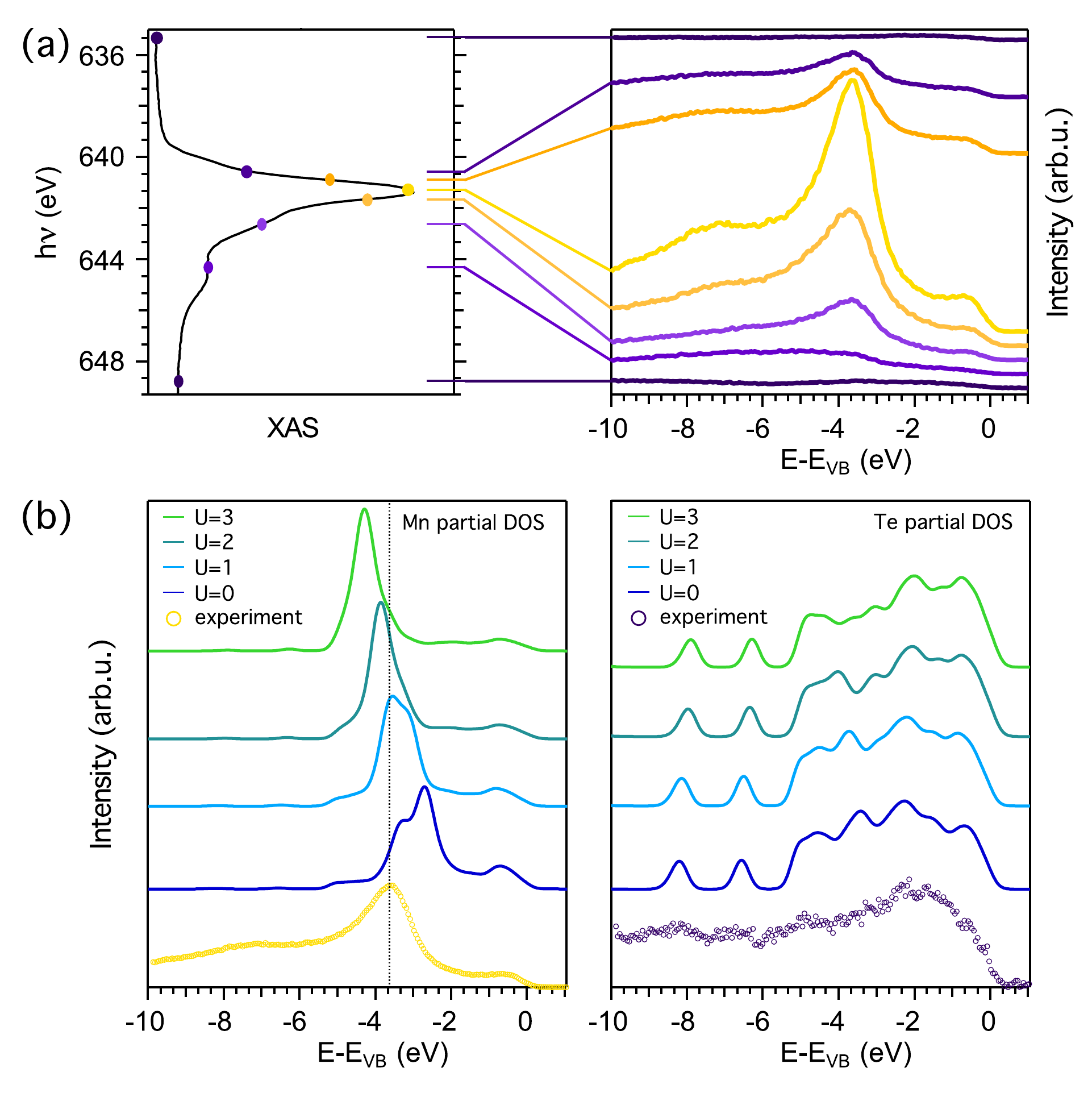}
\caption{\label{fig:fig3} Orbital hybridisation in the Mn$_{3}$Si$_{2}$Te$_{6}$ valence band. (a) Angle integrated, resonant photoemission (ResPES) of the valence band measured scanning the excitation energy across the Mn $L_3$ absorption edge. Coloured circles on the XAS spectra (\emph{left panel}) mark the photon energies at which the valence band photoemission measurements were performed (\emph{right panel}). The energy is referenced to the top of the valence band (E$_\mathrm{VB}$) and the VB spectra are vertically offset for clarity. (b) The Mn-(\emph{left}) and Te-(\emph{right}) partial density of states (P-DOS) as determined by DFT calculations as a function of $U$ and compared to the experimental data. The Mn partial density of states was experimentally evaluated as the difference between the on- and off-resonance photoemission spectra (measured at $h\nu=641.4$~eV and $h\nu=635.5$~eV, respectively) while the Te partial DOS was taken as the VB measured just before the Mn absorption edge (i.e., at $h\nu=635.5$~eV). Best agreement between the experimental and calculated data (see, e.g., dashed line as a guide to the eye) is found for the Mn partial DOS calculated at $U=1-2$ eV.}
\end{figure}

To address the role and extent of this implied covalency, we combine resonant photoemission spectroscopy (resPES) measurements of the valence band electronic structure with calculations from density-functional theory. Fig.~\ref{fig:fig3}(a) shows the valence band photoemission measured while scanning the photon energy across the Mn $L_3$ absorption edge. The data show a prominent Mn-derived state located at $\sim3.6$~eV binding energy, whose photoemitted intensity closely follows the $L_3$ XAS signal. This can thus be attributed as a nominally localised Mn component, which carries the majority of the spin moment in this system. Nonetheless, we find that there are increases in spectral weight through the Mn resonance not just for this peak, but more broadly for the entire valence band region. This implies that Mn becomes hybridised throughout the valence bands, as is evident from the Mn partial density of states (P-DOS) extracted as the difference of the PES measurements performed on- and off- resonance shown in Fig.~\ref{fig:fig3}(b). 

This directly points to a strong hybridisation between Mn and Te orbitals in this system, which in turn can be expected to weaken the electronic correlations in MST. Such correlations are expected to play a crucial role in the exchange-coupling among the Mn spins \cite{Nature2021Seo,PRB2017May,PRB2021Ni,PRB2021Liu},
and a quantification of the extent of electronic correlation {\it vs.} covalency in MST is strongly required. To this end, we compare the experimental Mn and Te P-DOS (the later extracted from off-resonant measurements) with the results of DFT+$U$ calculations in Fig.\ref{fig:fig3}(b). While the Te P-DOS only weakly changes with increasing Hubbard-$U$ parameter used in the calculation, the intense Mn-derived peak shifts rapidly away from the Fermi level, while its shape also becomes modified. The prominent peak in the on-resonance resPES data provides a robust experimental feedback, allowing  best matching the Hubbard-$U$ parameter for MST from comparison with our Mn P-DOS calculations. Consistency between the experimental measurements and theoretical calculations is found only for $U$ in the range of $1-2$~eV. This is substantially lower than typical $U$-values of about $4$~eV found for Mn$^{2+}$ oxides \cite{PRB2007Franchini} and even other chalcogenides such as MnSe$_2$ \cite{JPCM2021Xie} and MnTe$_2$ \cite{JAC2022Ma}, and lower than previously utilised for calculations of MST~\cite{PRB2017May,PRB2022Wang}. A deep physical interpretation of the $U$ value within the DFT+$U$ approach is not straightforward, due to the lack of formal correspondence between DFT+$U$ and the full many-body approach characteristic of the Hubbard model. Nonetheless, the small value of the Hubbard parameter here points to a significant reduction of correlations in MST, due to pronounced ligand hybridisation.

\begin{figure*}
\includegraphics[width=.8\textwidth]{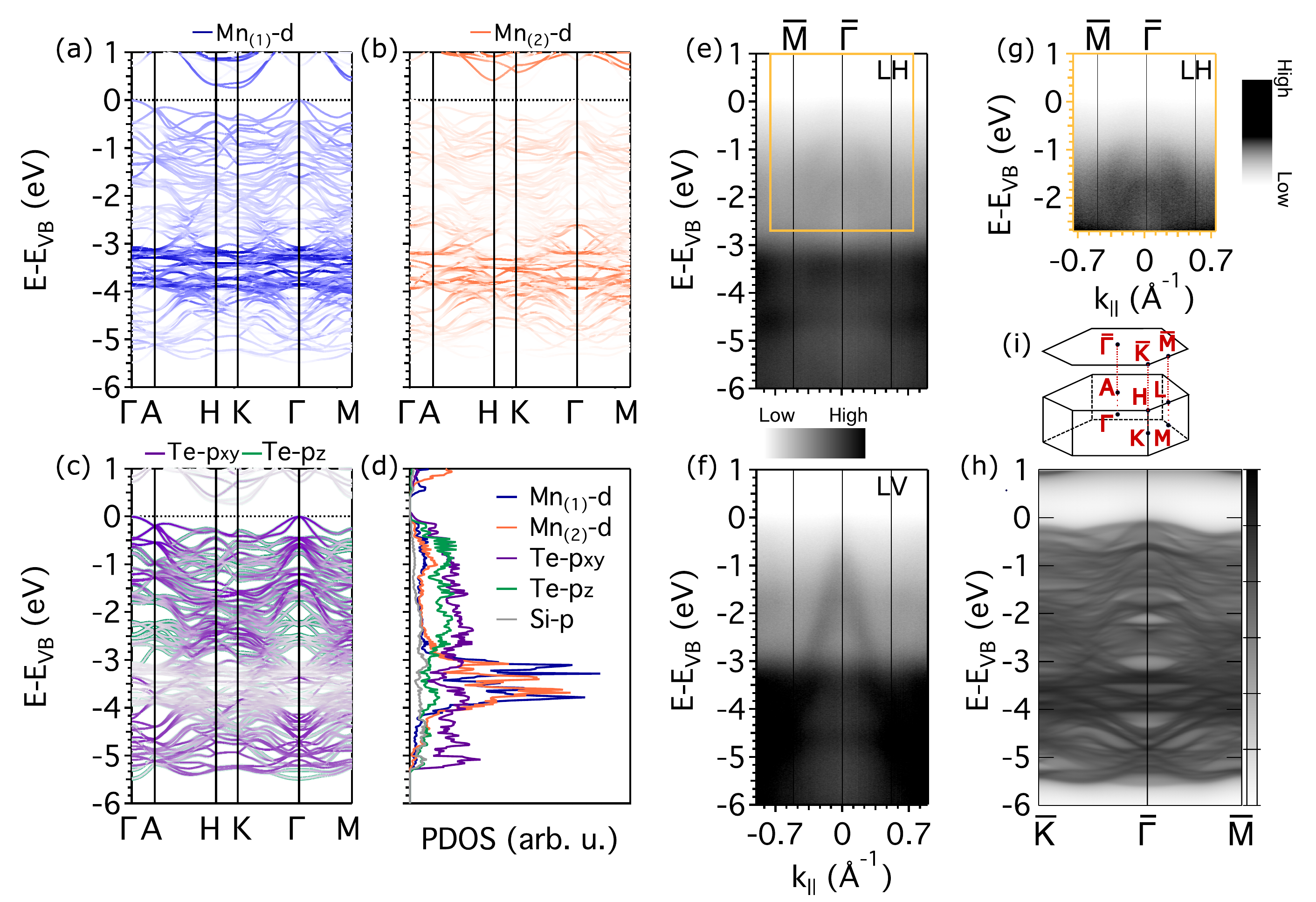}
\caption{\label{fig:fig4} Electronic structure of Mn$_{3}$Si$_{2}$Te$_{6}$. Calculated band dispersions along high symmetry directions, projected on the (a) Mn$_{(1)}$ and (b) Mn$_{(2)}$ $d$-orbitals and (c) the Te $p$-orbitals. (d) shows the corresponding partial DOS. Calculations were performed with $U=1$~eV. ARPES spectra acquired along the $\bar{\Gamma}$ - $\bar{\text{M}}$ at $58.5$~eV photon energy for both (e) horizontal (LH) and (f) vertical (LV) linearly polarized light, which show a strong dichroism particularly for the Te-derived states. (g) A magnified view of (e) with enhanced contrast, better highlighting the upper Te-bands. (h) Bulk states projected on the (001) surface Brillouin zone plane. (i) Bulk and surface-projected Brillouin zone marking the relevant high-symmetry points.}
\end{figure*}

In the occupied states, our calculations ($U=1$~eV, Fig.~\ref{fig:fig4}(a-d)) indicate a set of flat bands visible at an energy of $\sim\!3-4$~eV below the valence band top. These match well the peak in the Mn P-DOS visible in our res-PES measurements discussed above, as well as a corresponding non-dispersive feature visible in our measured dispersions from angle-resolved photoemission spectroscopy (ARPES, Fig.~\ref{fig:fig4}(e),(f)). Our orbitally-projected calculations indicate that these derive dominantly from the Mn states. However, consistent with our x-ray spectroscopic measurements, there is a strong hybridisation of the Mn states throughout the valence region. Interestingly, the two inequivalent Mn sites hybridise differently with the ligand valence states. Mn$_{(1)}$ (Fig.~\ref{fig:fig4}(a)), which sits within the MnSiTe$_3$ layer (Fig.~\ref{fig:fig1}(a)), has significant weight throughout the entire valence band bandwidth, while Mn$_{(2)}$, which sits in the interstitial sites between the layers, has a strongly localised Mn peak, exhibiting less intermixing with the Te $p$-orbitals across the rest of the valence band.

The Te states, meanwhile, contribute rather dispersive hole-like bands across a broad bandwidth of $\sim5$~eV (Fig.~\ref{fig:fig4}(c)). Our orbitally-resolved calculations indicate that the valence band maximum (VBM) is mostly contributed by in-plane Te $p_{xy}$-states and Mn$_{(1)}$, with wavefunctions thus mainly localised within the honeycomb layer. On the other hand, Te $p_z$ states provide a channel for inter-layer hybridization between the Mn$_{(1)}$ and Mn$_{(2)}$ ions. While many states are observed in our calculations due to the large unit cell of MST, the photoemission spectral weight enhances two such dispersive states (Fig.~\ref{fig:fig4}(e-g)), with the hybridized hole-like dispersion appearing rather broad in the ARPES measurements. This is consistent with them having a three-dimensional character, while the surface sensitivity intrinsic to ARPES, combined with the large $c$-axis lattice constant (see supplementary Fig.S1), leads to substantial $k_z$ broadening. This is also reflected in the surface-projected DFT calculations shown in Fig.~\ref{fig:fig4}(h). Aside from the surface-projected band features, broad background intensity is seen across the whole valence band energy range (up to $-6$~eV), broadly consistent with our experimental results. 

Finally, we note that there is a strong demarcation in the contribution of the Mn$_{(1)}$ and Mn$_{(2)}$ states to the conduction band electronic structure as can be readily seen in Fig.\ref{fig:fig4}(a,b). The lowest energy states are contributed almost entirely from the Mn$_{(1)}$ site, with some intermixing of the Te, while the Mn$_{(2)}$ states are located approximately $0.70-0.85$~eV higher in energy. The different chemical environments of Mn$_{(1)}$ and Mn$_{(2)}$ may again account for their different contributions to the conduction bands, with the lowest conduction band minimum found to be located almost entirely in the honeycomb layer.

\subsection{Exchange interactions}
The above comparison between DFT and our spectroscopic results strongly constrains the relevant $U$ parameter to relatively small values in MST. This, in turn, has a sizeable influence on the magnetic interactions and ordering tendencies in this system. To explore this, we have developed a magnetic model, where the exchange coupling constants have been estimated by mapping the {\it ab initio} total energies onto a classical Heisenberg model, that can be generally expressed as:
\begin{eqnarray}\label{eq:model}
H &=& \frac{1}{2}\sum_{ij} \bm S_i \cdot \mathcal{J}_{ij}\cdot \bm S_j +\sum_i\bm S_i \cdot \mathcal{A}_{i}\cdot \bm S_i
\end{eqnarray}
Exchange interactions between classical spins at sites $i,j$ are described here by the tensor $\mathcal{J}_{ij}$, including anisotropic effects, while the second term accounts for the magnetic single-ion anisotropy (SIA). The full exchange tensor can be decomposed into its isotropic part $J^{\mathrm{iso}}_{ij}=\frac{1}{3}\mathrm{Tr}{\mathcal J}_{ij}$, an antisymmetric term ${\mathcal J}^A_{ij}=\frac{1}{2}({\mathcal J}_{ij}-{\mathcal J^T}_{ij}$) and a symmetric term ${\mathcal J}^S_{ij}=\frac12({\mathcal J}_{ij}+{\mathcal J^T}_{ij})-J^{\mathrm{iso}}_{ij}\bm I$.

In order to account for the covalency in MST, we have considered up to fifth nearest-neighbor isotropic exchange interactions in our model. Consistent with previous studies, we label the first and second Mn$_{(1)}$-Mn$_{(2)}$ nearest-neighbour coupling as $J^{\mathrm{iso}}_1$ and $J^{\mathrm{iso}}_3$, while we consider two additional in-plane Mn$_{(1)}$-Mn$_{(1)}$ exchange interactions: $J^{\mathrm{iso}}_4,J^{\mathrm{iso}}_5$ (see Fig.~\ref{fig:fig1}(b)), as well as the next nearest-neighbour $J^{\mathrm{iso}}_2$\cite{PRB2017May,Nature2021Seo}. The exchange parameters have been extracted using the four-state energy mapping method\cite{PRB2011Xiang,Dalton2013Xiang}: this is a supercell approach that allows extracting the full exchange tensor describing the coupling between a selected pair of magnetic sites at a given distance, while the interaction with all other magnetic sites is canceled out by a tailored choice of four magnetic configurations (details can be found in the appendices of Ref. \cite{Dalton2013Xiang}). The four-state method also allows us to extract the single-ion anisotropy, {\it i.e.}, a site-dependent local quantity, instead of simply the total magnetic anisotropy energy. This is crucial for MST, comprising two inequivalent Mn sites that may in principle display different SIA. The anisotropic part of the exchange tensor of nearest-neighbor Mn$_{(1)}$-Mn$_{(2)}$ ($\mathcal{J}_1$) and Mn$_{(1)}$-Mn$_{(1)}$ ($\mathcal{J}_2$) were also considered: noticeably, a non-negligible Dzyaloshinksii-Moriya interaction is found between Mn$_{(1)}$-Mn$_{(2)}$ magnetic moments, with the Dzyaloshinskii vector parallel to the $c$-axis ($D_z$), while the Mn$_{(1)}$-Mn$_{(1)}$ exchange tensor only displays a symmetric anisotropic part. We emphasize that  the inclusion of {\em i}) the exchange coupling in tensorial form, {\em ii}) different single-ion anisotropies for the two  Mn atomic species, and {\em iii}) longer-ranged $J_4$ and $J_5$ exchange interactions, not considered in the literature, thereby improves our description with respect to previous models.  

In tables Tab. \ref{tab_magnetic_iso} and Tab. \ref{tab_magnetic_aniso} we list all the magnetic parameters estimated for $U=1$ eV and $U=2$ eV.
\begin{table}
\begin{tabular}{c|p{1cm}p{1cm}p{1cm}p{1cm}p{1cm}p{1cm}p{1cm}c}
     (meV)& $J^{\mathrm{iso}}_1$ &$J^{\mathrm{iso}}_2$ &$J^{\mathrm{iso}}_3$ &$J^{\mathrm{iso}}_4$ &$J^{\mathrm{iso}}_5$ & $A_c^{\mbox{\scriptsize Mn$_{(1)}$}}$ &$A_c^{\mbox{\scriptsize Mn$_{(2)}$}}$ &\\
    \hline
$U=1$ & {\centering 26.06} &{\centering 3.87} &{\centering 9.25} &{\centering 0.84} &{\centering 2.57} & 0.33 & 1.14\\
$U=2$ & {\centering 20.68} &{\centering 2.33} &{\centering 6.39} &{\centering 0.52} &{\centering 1.61} & 0.13 & 1.02\\
\end{tabular}
\caption{Estimated isotropic exchange interactions and single-ion anisotropies of MST using two different values of $U$. Within our definition of the Heisenberg model, a positive energy corresponds to an antiferromagnetic interaction. Fulfilling the symmetry properties of the system, SIA can be parametrized by a unique coupling constant $A_c=\mathcal{A}_{zz}-\mathcal{A}_{xx}$. All values are in units of meV, assuming classical spins of length $\vert S\vert=1$.}\label{tab_magnetic_iso}
\end{table}
\begin{table}
\begin{tabular}{c|p{1cm}p{1cm}p{1cm}p{1cm}p{1cm}p{1cm}p{1cm}c}
   (meV) & $J^S_{1xx}$ &$J^S_{1zz}$ &$D_{1z}$ &$J^S_{2xx}$ &$J^S_{2yy}$ &$J^S_{2zz}$ &$J^S_{2xz}$ &\\
    \hline
$U=1$ & {\centering -0.02} &{\centering 0.04} &{\centering 0.66} &{\centering -0.05} &{\centering 0.01} &{\centering 0.04} &{\centering 0.04} &\\
$U=2$ & {\centering -0.03} &{\centering 0.06} &{\centering 0.49} &{\centering -0.05} &{\centering 0.02} &{\centering 0.04} &{\centering 0.04} &\\
\end{tabular}
\caption{Estimated anisotropic exchange interactions of MST using two different values of $U$. Symmetry imposes $J_{1xx}=J_{1yy}$, while the cartesian components of tensor $J^S_2$ are given in a local reference frame with the $y$ axis perpendicular to the Mn$_{(1)}$-Mn$_{(1)}$ bond and the $z$ axis parallel to the crystallographic $c$ vector. The Dzyaloshinskii vector component is defined as $D_{1z}=(J_{1xy}-J_{1yx})/2$. All values are in units of meV, assuming classical spins of length $\vert S\vert=1$.}\label{tab_magnetic_aniso}
\end{table}
 Consistent with previous studies\cite{PRB2017May,Nature2021Seo}, all (isotropic) interactions are found to be antiferromagnetic, denoting a non-negligible magnetic frustration. On the other hand, our calculations also reveal sizeable longer-range interactions between Mn$_{(1)}$ magnetic moments within the honeycomb layers, with the third-nearest-neighbor exchange $J^{\mathrm{iso}}_5$ being of the same order of magnitude as $J_2^{\mathrm{iso}}$. This is likely due to both the presence of Si-Si dimers, which are located at the center of the Mn$_{(1)}$ hexagons where they can efficiently mediate the magnetic interactions, as well as to ligand contributions~\cite{PRB2022Riedl}.  While the (antiferromagnetic) Mn$_{(1)}$-Mn$_{(1)}$ interactions within the honeycomb layer do lead to magnetic frustration of the system, they are much smaller than the (antiferromagnetic) Mn$_{(1)}$-Mn$_{(2)}$ exchange coupling, so that a ferrimagnetic configuration with antiparallel Mn$_{(1)}$ and Mn$_{(2)}$ is expected to be the lowest energy state. 

The $J^{\mathrm{iso}}_4, J^{\mathrm{iso}}_5$ exchange interactions had not previously been estimated in Refs. \cite{PRB2017May,Nature2021Seo}, where total energies of different magnetic configurations defined within the crystallographic unit cell were mapped onto a classical Heisenberg model. We confirmed that such an approach cannot provide information on $J^{\mathrm{iso}}_4$, while $J^{\mathrm{iso}}_5$ interactions give rise to a spurious contribution to the nearest-neighbor Mn$_{(1)}$-Mn$_{(1)}$ exchange that, within this total-energy mapping procedure, would correspond to $J^{\mathrm{iso}}_2+J^{\mathrm{iso}}_5$. Taking into account the differences between the four-state and the total-energy mapping procedures, our results for $J^{\mathrm{iso}}_1, (J^{\mathrm{iso}}_2+J^{\mathrm{iso}}_5),J^{\mathrm{iso}}_3$ are in excellent agreement with previously reported estimates\cite{Nature2021Seo}. On the other hand, both SIA and anisotropic exchange interactions support the experimentally observed easy-plane magnetic anisotropy here. 

The magnetic anisotropy energy per Mn ion, defined as the energy difference $E_{\mbox{\scriptsize MAE}}=E_\perp-E_\parallel$ of a ferrimagnetic configuration perpendicular or parallel to the honeycomb ($ab$) layer, can be expressed from our model in Eq. \ref{eq:model} as:
\begin{eqnarray}
E_{\mbox{\scriptsize MAE}}&=&\frac{1}{3}\left(2A_c^{\mbox{\scriptsize Mn$_{(1)}$}}+A_c^{\mbox{\scriptsize Mn$_{(2)}$}}\right)+\frac{2}{3}\left(J_{1zz}-J_{1xx}\right)\nonumber\\
&&+\left[J_{2zz}-\frac{1}{2}\left(J_{2xx}+J_{2yy}\right) \right].
\end{eqnarray}
Using the parameters listed in Tables \ref{tab_magnetic_iso} and \ref{tab_magnetic_aniso}, we find a MAE of $\sim 0.70$ and $\sim 0.54$~meV/Mn for $U=1$~eV and $U=2$~eV, respectively. The dominant contribution to the MAE is provided by the Mn single-ion anisotropies. Here, the inequivalency of the Mn atoms is clearly reflected via distinct SIAs: both display an easy-plane character with the hard axis parallel to the crystal $c$ axis, but the SIA of Mn$_{(1)}$ is one order of magnitude smaller than that of Mn$_{(2)}$. Interestingly, we found that roughly 15\% of the magnetic anisotropy energy is contributed by anisotropic exchange interactions. In agreement with previous studies~\cite{PRB2017May,Nature2021Seo}, we find that the dominant Mn$_{(1)}$-Mn$_{(2)}$ antiferromagnetic interactions favour a ferrimagnetic ground state comprising ferromagnetic Mn$_{(1)}$ and Mn$_{(2)}$ layers antiferromagnetically aligned. 

\begin{figure*}
\includegraphics[width=.9\textwidth]{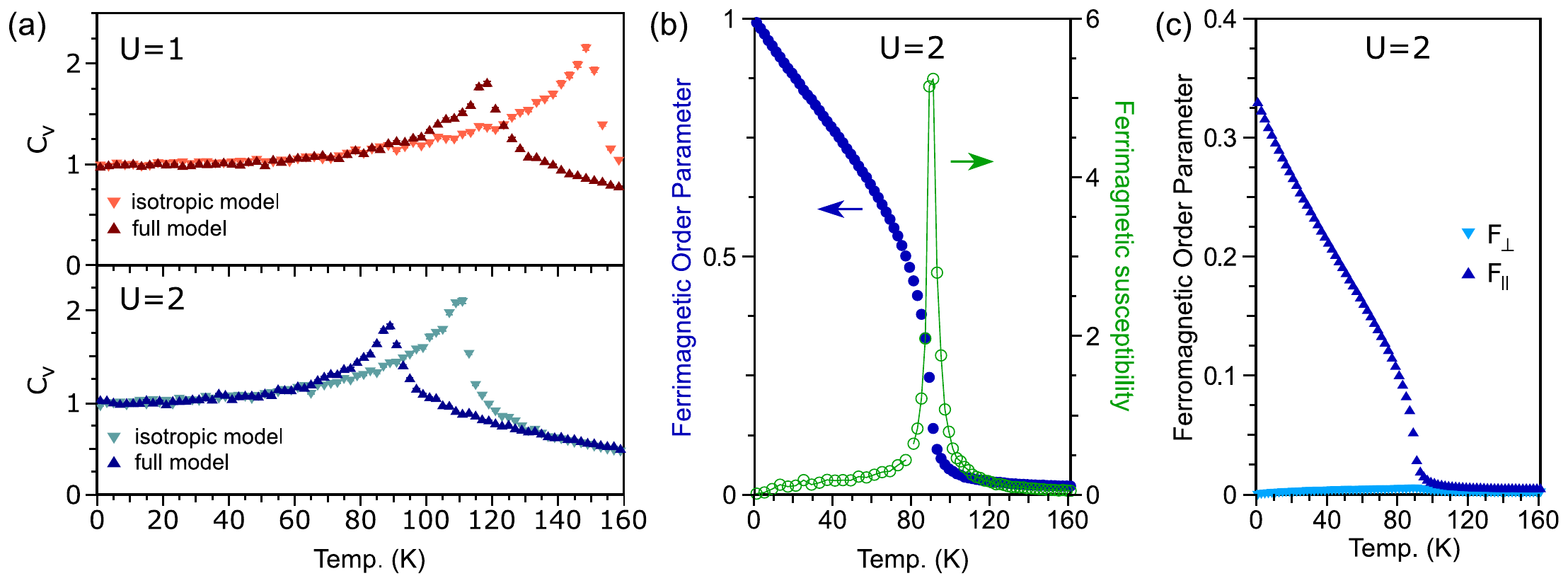}
\caption{\label{fig:figMC} 
Monte Carlo simulations. (a) Specific heat of the full model (dark symbols) with parameters obtained from DFT+$U$ calculations for $U=1,2$, compared with the specific heat calculated on a simplified model with isotropic $J_1^{\mathrm{iso}}, (J_2^{\mathrm{iso}}+J_5^{\mathrm{iso}}),J_3^{\mathrm{iso}}$ (light symbols). The stronger magnetic frustration of the full model reduces the critical temperature $T_\mathrm{C}$, signalled by the peak in $C_v$, by roughly 20\% with respect to the simplified model. (b) Temperature evolution of the ferrimagnetic order parameter and its associated susceptibility which displays a sharp peak at the transition (only $U=2$~eV results are shown; the same trend was observed with the other set of parameters) and signalling the transition to a ferrimagnetic state. (c) The in-plane $F_\parallel=\sqrt(F_x^2+F_y^2)$ and out-of-plane $F_\perp=\vert F_z \vert$ components of the ferromagnetic order parameter are shown as a function of temperature, confirming the easy-plane character of the magnetically ordered phase. }
\end{figure*}

We estimate $T_\mathrm{C}$ from our calculated exchange couplings using Monte Carlo simulations (Fig.~\ref{fig:figMC}). The transition temperature extracted from the temperature dependence of the calculated specific heat per spin, $C_v$, is estimated to be $T_c\simeq 89$~K  and $T_c\simeq 118$~K from our calculations taking $U=2$ eV and $U=1$ eV, respectively. These are in good qualitative agreement with the experimentally determined $T_C=75-79$~K, further supporting the identified range of $U=1-2$~eV as appropriate for MST, and highlighting the crucial role of magnetic anisotropy, which leads to a significantly reduced $T_\mathrm{C}$ in our calculations in comparison to an isotropic model. Our findings from analysis of the calculated specific heat are further validated by direct extraction of the ferrimagnetic order parameter (Fig.~\ref{fig:figMC}(b), see Methods) which, together with the abrupt change of its relative susceptibility, shows the ferrimagnetic state onset. Finally, we note that the dependence of the ferromagnetic order parameter with temperature (Fig.~\ref{fig:figMC}(c), see Methods) indicates the magnetic easy-plane nature of the system, further validating our discussions above.

\section{Conclusions}
Our results demonstrate an integrated approach to predicting and understanding the interactions governing long-range magnetic order in the layered ferrimagnet Mn$_3$Si$_2$Te$_6$. Our approach allows characterising the importance of covalency from ligand-metal hybridisation, which we find to be significant in this system, weakening the electronic correlations. Using these spectroscopic results to constrain our calculations, we demonstrate how combined first-principles and Monte Carlo methods can be used to accurately predict exchange interactions and magnetic anisotropies, elucidating their role in stabilising MST's novel ferrimagnetic order and predicting its ordering temperature within $\sim\!15$~K of the experimental value. Our results reveal a key role of covalency in ruling these properties in MST, which we expect to be key to understanding magnetism across the emerging class of 2D and layered magnets.

\section*{Acknowledgments:} We gratefully acknowledge support from The Leverhulme Trust via Grant No.~RL-2016-006 and the European Research Council (through the QUESTDO project, 714193). P.B. and S.P. acknowledge financial support from the Italian Ministry for Research and Education through PRIN-2017 projects ‘Tuning and understanding Quantum phases in 2D materials—Quantum 2D’ (IT-MIUR grant No. 2017Z8TS5B) and ‘TWEET: Towards ferroelectricity in two dimensions’ (IT-MIUR grant No. 2017YCTB59), respectively. MCH, DM and GB acknowledge financial support by the UK Engineering and Physical Sciences Research Council through grant EP/T005963/1. We thank the Elettra synchrotron for access to the APE-HE beamline under proposal number 20195300. We thank Diamond Light Source for beamtime on the I10 beamline under proposal number MM28727-1. The research leading to this result has been supported by the project CALIPSOplus under Grant Agreement 730872 from the EU Framework Programme for Research and Innovation HORIZON 2020. G.V., V.P., D.D. and F.M. acknowledge financial support from the Nanoscience Foundry and Fine Analysis (NFFA-MUR Italy Progetti Internazionali) project (www.trieste.NFFA.eu). For the purpose of open access, the authors have applied a Creative Commons Attribution (CC BY) licence to any Author Accepted Manuscript version arising. The research data supporting this publication can be accessed at [[DOI TO BE INSERTED]].

\bibliography{thebibliography.bib}

\clearpage
\section{Supplementary Information:}

\renewcommand{\thefigure}{S\arabic{figure}}   
\renewcommand{\figurename}{Supplementary Figure}

\begin{figure}[!h]
\includegraphics[width=0.6\columnwidth]{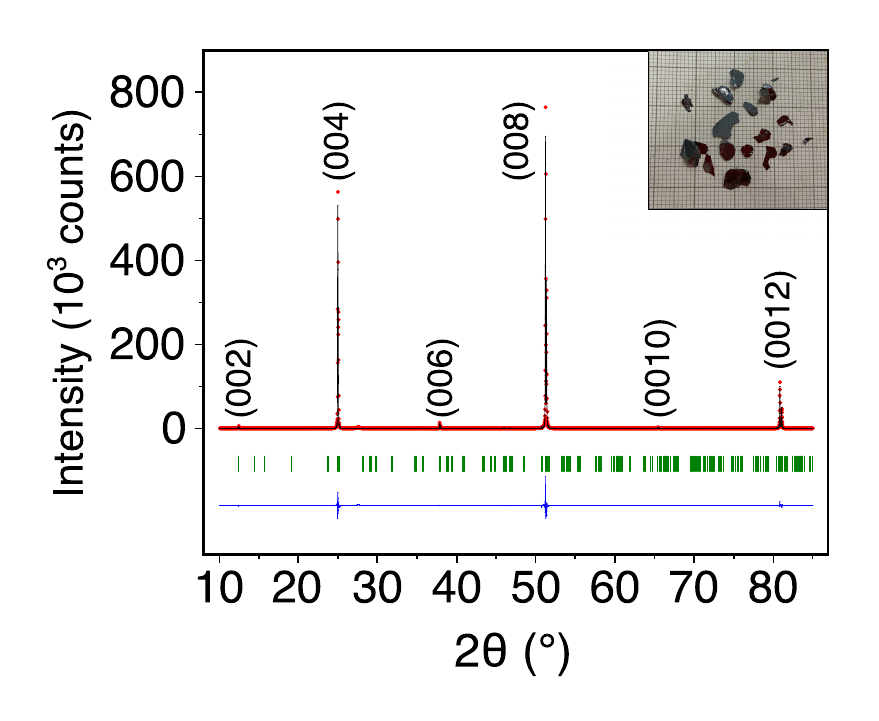}
\caption{\label{fig:figS2} X-ray Diffraction pattern of Mn$_{3}$Si$_{2}$Te$_{6}$ single crystal. The pattern shows the (00\emph{l}) reflections. The inset shows typical Mn$_{3}$Si$_{2}$Te$_{6}$ single crystals.}
\end{figure}

\begin{figure*}[!t]
\includegraphics[width=0.9\textwidth]{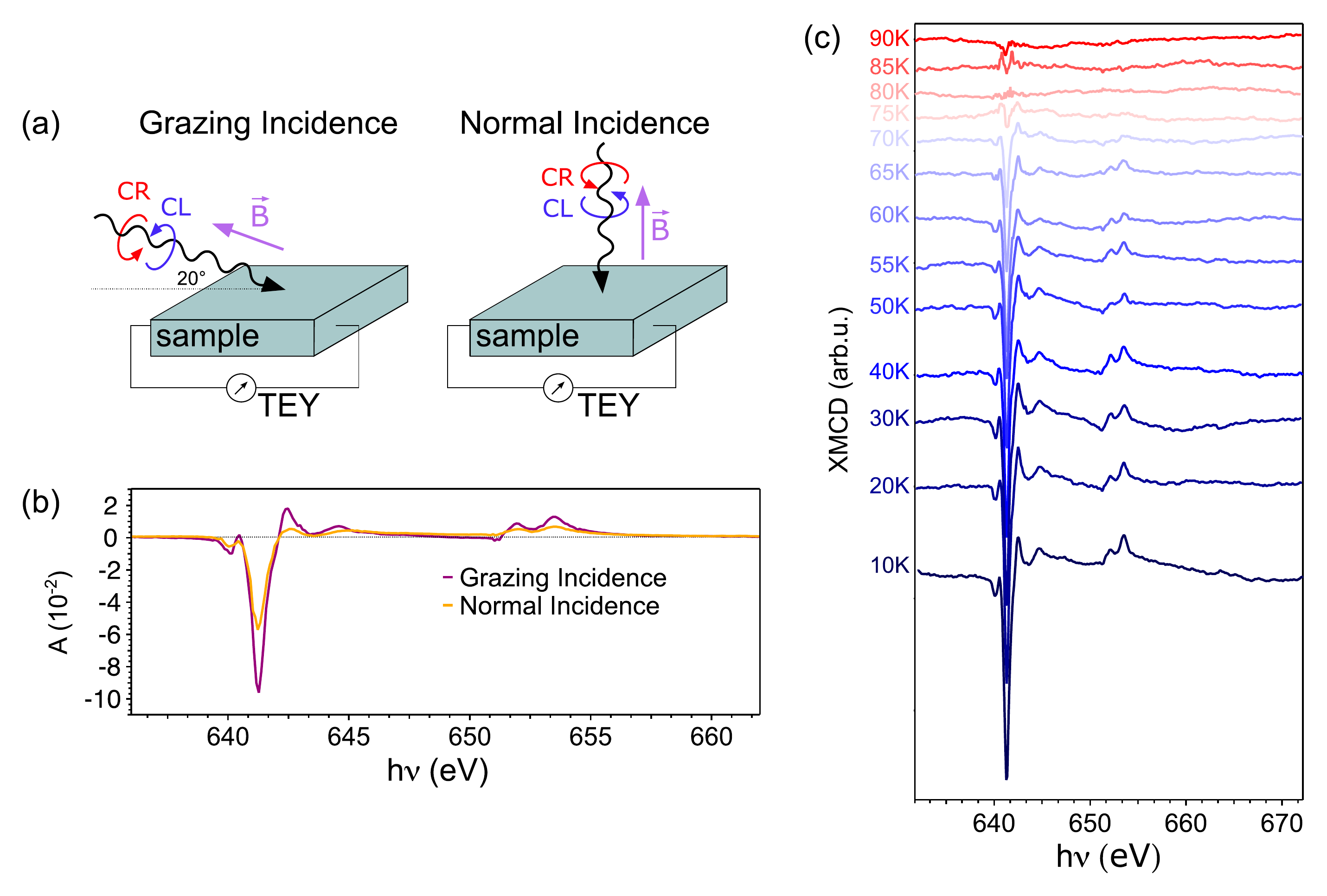}
\caption{\label{fig:figS3} (a) Experimental geometries for X-ray Magnetic Circular Dichroism (XMCD) measurements. In grazing incidence geometry (\emph{left}) circularly polarized light impinges on the sample at $20^\circ$ from the sample surface (i.e., the $ab$-plane), while the normal incidence geometry is achieved when the photon beam is aligned perpendicularly to the $ab$-plane. In both geometries, the magnetic field is collinear with the incident photon beam and the signal is acquired in total electron yield (TEY) detection by measuring the drain current of the sample. (b) XMCD asymmetry signal measured in the two different geometries with $-1.4$~T applied field and the sample temperature kept at $10$~K. The XMCD signal in grazing incidence being approximately twice larger than for the normal geometry identifies $ab$-plane as the easy magnetization plane and it is in good agreement with the bulk magnetic properties measured by SQUID (Fig.1(c), (d) in the main text). (c) Remanent XMCD spectra series of the Mn $L_{2,3}$ absorption edge measured in the Mn$_3$Si$_2$Te$_6$ $ab$-plane at different temperatures across the ferrimagnetic transition (i.e., between $10$~K and $90$~K) showing the onsetting of the long range magnetic order below $75$~K. Spectra are offset along the vertical axis for clarity.}
\end{figure*}

\end{document}